\documentclass[preprint,12pt]{aastex}

\shortauthors{Kobayashi and Kamaya}
\shorttitle{SPECTRA FROM FORMING REGION OF THE FIRST GALAXIES}

\begin{document}

\title{SPECTRA FROM FORMING REGION OF THE FIRST GALAXIES : THE EFFECT OF
ASPHERICAL DECELERATION}

\author{Masakazu AR. Kobayashi and Hideyuki Kamaya}

\affil{Department of Astronomy, School of Science, Kyoto University,
Sakyo-ku, Kyoto 606-8502, JAPAN}

\email{TO: MARK: kobayasi@kusastro.kyoto-u.ac.jp}

\begin{abstract}

Ly$\alpha$ line emission from the Loeb-Rybicki (LR) halo, which is the
expanding HI IGM (intergalactic medium) around the first star clusters
and the ionized interstellar medium, is investigated by solving a
radiative transfer problem.  
While the initial scattering optical depth is $\sim 10^5$ for the
Ly$\alpha$ photons, most of the Ly$\alpha$ photons can escape when the
cumulative frequency-shift due to the expansion of the HI IGM becomes
significantly large.  
The current paper improves upon previous treatments of the scattering
processes and the opacity for the Ly$\alpha$ transfer.  
Confirming the previous results of the LR halo, we investigate the
effect of the aspherical expansion of the IGM.  
The asphericity is hypothesized to follow the initial stage of the
gravitational deceleration to form the large scale filamentary structure
of the Universe.  
According to our results, the effect of the asphericity lets the peak
wavelength of the line profile shift to longer wavelengths and the FWHM
of the profile become wider than those of the spherically expanding
model.  
To detect these features is meaningful if we are interested in the
initial evolution of the large scale structure, since they reflect the
dynamical properties of the IGM at that time. 
Furthermore, given the recent discovery of the high redshift
cosmological reionization, we briefly comment on the effects of the
redshift and the cosmological parameters on the line profile.

\end{abstract}


\section{INTRODUCTION}

The epoch of cosmic reionization is one of the most important epochs to
understand in the investigation of the history of the universe, because
it contains the imprint of important information on cosmic structure
formation and evolution. 
This is because the reionization of the intergalactic
medium (IGM) occurs promptly (e.g. Gnedin \& Ostriker 1997) and the
cosmic ionizing background radiation can penetrate the dense HI
condensations of virialized objects (Miralda-Escud\'{e}, Haehnelt, \&
Rees 2000; Barkana \& Loeb 1999).

The epoch of the reionization observationally relates to the
Gunn-Peterson trough (Gunn \& Peterson 1965).  The significantly ionized
IGM in the observed spectra of QSOs at $z < 5.6$ (Dey et al. 1998; Hu,
Cowie, \& McMahon 1998; Spinrad et al. 1998; Weymann et al. 1998)
implies that most of the intergalactic hydrogen in the Universe has been
reionized before $z \sim 5.6$.  Some standard cosmological models
predict that reionization takes place around a redshift of $z \sim 6$.
However, according to the results of the WMAP project, Kogut et al. (2003)
may have discovered the high redshift reionization of the Universe ($z\sim
20$).  
To resolve this discrepancy, the direct detection of the emission from
the first objects is very important.  
This is because that emission reflects the physical and chemical
properties of the first star clusters (Shchekinov 1991; Kamaya \& Silk
2002) and/or protogalaxies. 
Once we know the origin of the first emission, the ionization rate at
that time is investigated.  
Therefore, new detailed studies and confirmation of the previous studies
of the first emission in another way are important for the
subsequent progress of the research fields.

It is interesting to detect just the first Ly$\alpha$ photons since the
star formation rate and/or the ionization structure of the interstellar
medium are examined from that intensity.  
Using standard cosmological models, cosmic recombination takes place at
$z\sim 10^3$.  
After that epoch, the Universe becomes predominantly neutral.
When the first galaxies and/or star clusters form, they are surrounded
by the neutral IGM.  
One might naively expect the damping wing of the Ly$\alpha$ trough to
eliminate any trace of a Ly$\alpha$ emission line in the observed source
spectra.  
However, the Ly$\alpha$ photons are scattered and diffused redward in
frequency owing to the Hubble expansion of the surrounding HI IGM
(Loeb \& Rybicki 1999 ; hereafter LR99).  
When the cumulative frequency-shift becomes significantly large, they
can escape and travel freely toward the observer.

Although we investigate a similar problem to that in LR99, the
radiative transfer model is improved in the two ways : (1) Formulation
of the geometrical scattering (i.e. test photons scatter spatially three
dimension); and (2) Formulation of the Ly$\alpha$ opacity.  
Thanks to these two improvements and the confirmation of the previous
studies, we can study more complex physical conditions in the expanding
HI gas.  
Since the large scale structure (LSS) forms filaments, this suggests the
initial stage of the gravitational deceleration for the LSS favors an
asymmetrical expansion of the HI halo.   
Indeed, the inhomogeneity of the self-gravitating system can grow in a
free-fall time (e.g. Lin, Mestel, \& Shu 1965).  
As long as the expansion time-scale is comparable to the free-fall time
scale, the asymmetrical deceleration of the IGM occurs in the time-scale
of the Hubble expansion, although the initial asymmetry is obviously
very weak. 
As the first and the simplest hypothesis to investigate the effect of the
inhomogeneity via the gravitational deceleration for the LSS formation,
we adopt the aspherical expansion model of HI IGM.  
The physical implication drawn from the current analysis will suggest to
us further research projects for the first Ly$\alpha$.  
Since the lumpiness and/or small-scale irregularity may be important for
a realistic treatment of the first Ly$\alpha$, we will investigate these
effects in subsequent papers. 

This paper is organized as follows: In \S 2 we describe the formalism of
the Ly$\alpha$ line transfer problem commenting on the differences
between our approach and LR99.
In \S3 we outline our Monte Carlo simulation method.
In \S4 we discuss our numerical results including the effects of a more
complete representation of the Ly$\alpha$ opacity, the effect of
asphericity in the expanding IGM, and the sensitivity of our results to
redshift and cosmological parameters.
In \S 5 we summarize our study.

\section{FORMULATION}

We formulate the problem according to LR99, while the HI IGM is
hypothesized to expand aspherically.
In general, the exactly spherical expansion is not expected as long as
the IGM decelerates to form LSS. 
This is recognized because the free-fall time, $t_{ff}$, is comparable
to the Hubble expansion time, $t_h$, as explained in \S 1.
Then, to find the effect of the asymmetry, we shall set the most simple
assumption to the recession velocity, {\it spheroidal expansion}:
$v = H_s\left[\varepsilon^2 \left(x^2+y^2\right)+z^2\right]^{1/2}$,
where $v$ is recession velocity, $H_s$ is the Hubble expansion rate at
the source redshift of $z_s$, and $\varepsilon$ is {\it ellipticity} or
decreasing rate of the recession velocity to the pure Hubble expansion.
We shall estimate the allowed order of $\varepsilon$. 
The ratio of the two time-scales is denoted as
\begin{equation}
 \frac{t_{ff}}{t_h}\sim \frac{\pi}{2}\Omega_M^{-1/2}\delta^{-1/2}.
\end{equation}
Here, $\delta$ is the ratio of the density of the halo to the mean
density of the Universe at $z$. 
Of course, this ratio must not be different from unity very much
as long as IGM expands.
Since the density of the aspherically decelerating HI halo is 
larger than that of the spherically expanding halo, we find $\delta =
1/\varepsilon$ with mass conservation.
Fixing the ratio of Eq.(1) to be unity and 
postulating $\delta \sim 6$, for example, we estimate $\varepsilon 
\sim 0.2$ for $\Omega_M = 0.4$.
Thus, we recognize that order of $\varepsilon$ is about 0.1.
The case of $\varepsilon = 0.1$ is considered as an extreme case.
Obviously, $\varepsilon = 1$ corresponds to LR99.

Although $\varepsilon$ should approach to unity differentially from
small scale to large scale, we assign $\varepsilon$ a single value for
simplicity.  
Fortunately, the size of the concerned region, $\sim r_{\ast}$, which is
a characteristic distance from the Ly$\alpha$ photon source as defined
later, is $\sim$ Mpc as estimated in LR99.
That is, $r_{\ast}$ corresponds to the size of groups of galaxies and
the region at $r_{\ast}$ can be regarded as decelerating region
although the allowed deceleration should be small (but not zero).  
Expansion law of the spheroid approaches to that of the
spherical model at larger scale than $r_{\ast}$.
Properties of test photons are also determined at $\sim r_{\ast}$.
Hence, the single value hypothesis for $\varepsilon$ can present the
physical insight into the effect of the deceleration for the LR halo. 
In realistic condition, by the way, $\varepsilon$ inside $r_{\ast}$
should decrease toward the center differentially.
Then, our finding characteristic effect of the deceleration may be
enhanced because the number of scattering can increase owing to the
deceleration as shown in this paper.

Let $I = I(\nu ,{\bf \Omega}, {\bf r})$ be the specific intensity (in
photons ${\rm cm^{-2}\ s^{-1}\ sr^{-1}\ Hz^{-1}}$), where $\nu$ is
comoving frequency, ${\bf r} = \left(x, y, z\right)$ is the radial
vector from the source galaxy, and ${\bf \Omega}$ is the scattering
direction. 
The spatial components of the direction cosines of ${\bf \Omega}$ are 
$\left(u, v, w\right)$, respectively.
With the assumption of isotropic coherent scattering, the comoving
transfer equation for a resonance line is given by
\begin{equation}
 {\bf \Omega} \cdot \nabla I+ \alpha \psi 
  \left(\varepsilon, w\right) \frac{\partial I}{\partial \nu} 
  = \chi_{\nu} \left(J - I\right)+S . {\label{eq1}}  
\end{equation}
Here, $\nu$ is the shift of the frequency $\nu_{\alpha} - \nu_{\rm
photon}$ where $\nu_{\alpha}$ is the resonant frequency and  $\nu_{\rm
photon}$ is the photon frequency; $\chi_{\nu}$ is the scattering opacity
at $\nu$ (eq. (4) in LR99); $J=(1/2)\int_{-1}^1 I d \mu $ is the mean
intensity; $S$ is the emission function (in photons ${\rm cm^{-3}\
s^{-1}\ sr^{-1}\ Hz^{-1}}$, eq. (3) in LR99); $\alpha \equiv
H_s\nu_{\alpha}/c$ is the increasing rate of the frequency shift per
unit distance in the spherically expanding IGM (eq. (2) in LR99); and 
\begin{equation}
\psi \left(\varepsilon, w\right) 
 = \left[\varepsilon^2 + \left(1 - \varepsilon^2\right)w^2\right]^{1/2} 
\end{equation}
represents the effect of aspherical expansion.

In LR99 the scattering opacity $\chi_{\nu}$ was approximated as
$\chi_{\nu}\sim \beta /\nu^2$.
In the current paper, we formulate the opacity in the following more
complete form:
\begin{eqnarray}
 \chi_{\nu} &=& \frac{\beta / \varepsilon}{\Lambda_{\alpha} ^2/16\pi^2}
  ~~{\rm for }~~ \nu^2 \ll \Lambda_{\alpha}^2/16\pi^2, {\label{form1}}\\
 &=& \frac{\beta / \varepsilon}{\nu^2+\Lambda_{\alpha}^2/16\pi^2}
  ~~{\rm for }~~ \nu^2 \sim \Lambda_{\alpha}^2/16\pi^2, 
  {\label{form2}}\\
 &=& \frac{\beta / \varepsilon}{\nu^2}
  ~~{\rm for }~~ \Lambda_{\alpha}^2/16\pi^2 \ll \nu^2\ \& \ \nu \ll 
  \nu_{\alpha},
  {\label{form3}}\\
 &=& \left(1-\frac{\nu}{\nu_{\alpha}}\right)^4\frac{\beta / \varepsilon}
  {\nu^2}~~{\rm for }~~ \nu \sim \nu_{\alpha}, 
  {\label{form4}}\\
 &=& 0  ~~{\rm for }~~  \nu > \nu_{\alpha}, {\label{form5}}
\end{eqnarray}
where $\nu_{\alpha}=2.47\times 10^{15}\ {\rm Hz}$ is the Ly$\alpha$
frequency, $\Lambda_{\alpha} = (8\pi^2e^2f_{\alpha} / 3m_ec
\lambda_{\alpha}^2) = 6.26\times 10^8\ {\rm s^{-1}}$ is the rate of
spontaneous radiative decay from $2p$ to $1s$ energy levels of hydrogen,
$f_{\alpha}=0.4162$ is the oscillator strength, $\lambda_{\alpha} =
1216\ {\rm \AA}$ is the wavelength of the Ly$\alpha$ line, and $\beta$
is $1.5 \Omega _{\rm b} h_0^2 (1+z_{\rm s})^3$ cm$^{-1}$ Hz$^2$ 
(eq. (5) in LR99). In the initial scattering phase,
Eq.(4) is important since the frequency shift can be smaller
than $\Lambda_{\alpha}/4\pi$. When the photon diffuses and
the frequency shift is comparable to $\nu_{\alpha}$, 
Eq.(7) must be adopted as $\chi_{\nu}$. 

We normalize the frequency shift and radius by convenient quantities
like LR99.  
An appropriate frequency scale, $\nu_{\ast}$, is estimated from
$\tau_{\nu} \sim 1$. 
Here, $\tau_{\nu}$ is the Ly$\alpha$ optical depth from the source to
the observer, and is defined as  
\begin{equation}
 \tau_{\nu}=\int_0^{\infty}\chi_{\nu+\alpha \psi r}dr.
  {\label{dfnstar}}
\end{equation}
The characteristic frequency shift of $\nu_{\ast}$ is measured 
along the direction of $z$ (i.e. $w = 1$).
Eq. ({\ref{dfnstar}}) with Eq.(7) can be described as
\begin{equation}
 \int^{l_0}_0\left(1-\frac{\nu_{\ast}+\alpha r}{\nu_{\alpha}}\right)^4
  \frac{\beta /\varepsilon}{\left(\nu_{\ast}+\alpha r\right)^2} dr =1,  
  {\label{dfnstar2}}
\end{equation}
where $l_0 \equiv (\nu_{\alpha} - \nu_{\ast})/\alpha$ 
is a characteristic distance for the photon frequency to reach zero.
We note that $\nu_{\ast}$ always depends on $\beta$ and $\varepsilon$.
The appropriate radius, where the frequency-shift due to the Hubble
expansion produces the frequency shift of $\nu_{\ast} $, is given by
$r_{\ast} = \nu_{\ast}/\alpha$. 
For example, with $\varepsilon = 1$, we find $\nu_{\ast} = 9.64 \times
10^{12}\ {\rm Hz}$ and $r_{\ast}=0.77$ Mpc at $z_s=10$ with a set of
cosmological parameters of $\Omega_b = 0.05,\ \Omega_M =0.4,\
\Omega_{\Lambda} = 0.6,$ and $h_0 = 0.65$. 
\footnote{We adopt these cosmological parameters because of
the comparison to LR99.  
Although the recent results of WMAP present slightly different
cosmological parameters (Spergel et al. 2003), fortunately our results
are not altered very much as shown in \S 4.4.}
The normalized frequency shift $\tilde{\nu}$ and distance $\tilde{r}$
are then defined to be $\tilde{\nu}\equiv \nu /\nu_{\ast}$ and
$\tilde{r}\equiv r/r_{\ast}$.

\begin{figure}[htbp]
   \begin{center}
    \includegraphics{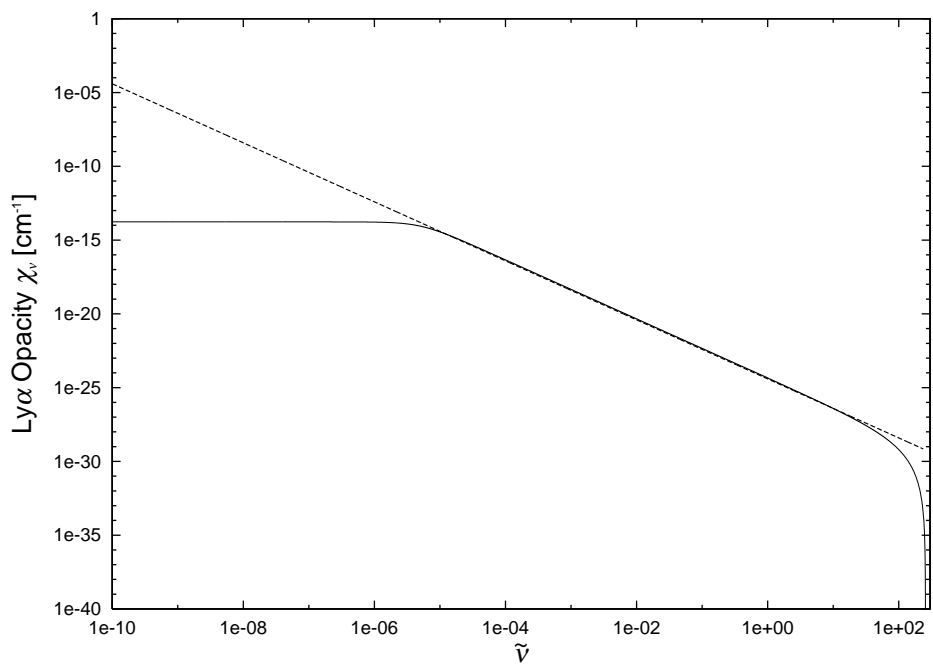}
    \caption{Ly$\alpha$ opacities are presented.
    Solid line represents the fully formulated opacity [Eq.(4) in LR99],
    and dashed line does the approximated opacity in LR99.}
    \label{fig-kai_nu_comp}
   \end{center}
  \end{figure}

For the reader's convenience, we present the Lyman$\alpha$ opacity
$\chi_{\nu}$ ${\rm cm^{-1}}$ in Fig.{\ref{fig-kai_nu_comp}} as a
function of $\tilde{\nu}$.  
The solid line denotes our $\chi_{\nu}$, and the dashed line denotes
the previously adopted one from LR99.
According to the definition of $\tilde{\nu}$, the photon frequency $\nu_{\rm
photon}$ becomes small as $\tilde{\nu}$ increases and reaches
zero at $\tilde{\nu} = \nu_{\alpha}/\nu_{\ast}\sim 2.56\times 10^2$. 
At $\tilde{\nu} = 0$, the frequency of the photon is the Ly$\alpha$
frequency $\nu_{\alpha}$.  
The differences depicted in the figure {\ref{fig-kai_nu_comp}} are
reflected in our results, while they are usually small.

 \section{Algorithm of the Calculation}

LR99 solved a dimensionless transfer equation in the diffusion regime,
which is valid in the region $\tilde{\nu}\gg \tilde{r}$.
They also studied the numerical solution of the transfer equation using
a Monte Carlo technique and found that it agreed with the diffusion
equation in the low frequency shift region.
We also solve the transfer equation numerically by using a Monte Carlo
method very similar to LR99.
The number of test photons is $10^8$, the same as in LR99, and permits
us to compare our results directly to theirs.
During the calculation two photon decay is always neglected since it is
difficult to destroy the Lyman $\alpha$ photons transferred in the
surrounding HI.
This is because the event rate of two photon decay is determined by the
process of a transition between $2s$ - $2p$ owing to the radiative
collision with an electron (Breit \& Teller 1940; Spitzer \& Greenstein
1951). 
As long as the relic electron number density is very small, the event
rate of two photon decay should be negligible.  

LR99 adopted the initial shift method in the first step of each
simulation. 
That is, the initial shift of $\tilde{\nu}$ is 0.01 and $\tilde{\bf r}$
is determined stochastically by means of the diffusion solution of
LR99. 
We have confirmed independently that the initial shift method is very
reasonable.  
This is because the initial optical depth is always much larger than
unity. 
Hence we also adopt that method in all first time steps of the
simulations. 

 \begin{enumerate}
  \item 3D scattering \\
LR99 assumed isotropic, coherent scattering through a spherically
symmetric, neutral medium such that the scattering could be described by
a single angular parameter $\mu = \cos{\theta}$ relative to the radius
$r$. 
We generalize to three dimensional scattering.
The scattering processes in 3D coordinates is described by two parameters
of $\mu =2R_1-1$ and $ \phi = \pi \left(2R_2-1\right)$. 
Here, $\phi$ is the change in azimuth, and $R_1$ and $R_2$ are random
numbers on the interval of (0, 1). 
Defining $\mu $ as $\cos{\alpha}$, 
we measure $\alpha$ in the plane determined by the old and new photon
paths.

The direction cosines of  $u_{i+1},v_{i+1},w_{i+1}$ after the scattering 
are determined by using the direction cosines of the incident photon,
$u_i, v_i, w_i$:  
\begin{eqnarray}
u_{i+1}&=&\sin{\alpha}\left(\cos{\phi}w_iu_i-\sin{\phi}v_i\right)
 /\left(1-w_i^2\right)^{1/2}+\cos{\alpha}u_i  ,  \nonumber  \\
v_{i+1}&=&\sin{\alpha}\left(\cos{\phi}w_iv_i+\sin{\phi}u_i\right)
 /\left(1-w_i^2\right)^{1/2}+\cos{\alpha}v_i {\label{eq-def_new_dcos}
 } , \\
w_{i+1}&=&-\sin{\alpha}\cos{\phi}\left(1-w_i^2\right)^{1/2}+\cos{\alpha}
  w_i.\nonumber
\end{eqnarray}
As $|w_i|$ approaches unity, according to Witt (1977), the set of equations
({\ref{eq-def_new_dcos}}) breaks down and is replaced by
\begin{equation}
u_{i+1}=\sin{\alpha}\cos{\phi},\ v_{i+1}=\sin{\alpha}\sin{\phi},\ 
w_{i+1}=\cos{\alpha}w_i\quad {\rm for}\ 1-|w_i|\le 10^{-10}   . 
\end{equation}

  \item Opacity selection \\
LR99 used a single approximate form for the opacity as a function of
frequency (i.e. our equation ({\ref{form3}}) for $\varepsilon = 1$). 
We use the more complete form for the opacity by Equations
({\ref{form1}}-{\ref{form5}}).
Selecting the appropriate opacity in Eqs.({\ref{form1}}-{\ref{form5}})
according to $\tilde{\nu}$, we evaluate the path length along
the ray, $\tilde{l}$, and  the maximum optical depth: $ \tau_{\rm max} 
\equiv \int_{0}^{\infty}\chi_{\nu+\alpha \psi l}dl$.
Once the optical depth $\tau$ for a test photon becomes larger than the
maximum optical depth $\tau_{\rm max}$, we say that it escapes from the
LR halo. 
Here, $\tau$ is determined as $-{\rm ln}R_{\rm sca}$ where $R_{\rm sca}$
is a random number between zero and one. 

  \item Aspherical expansion \\
LR99 assumed that the neutral IGM surrounding the Ly$\alpha$ point
source undergoes uniform Hubble expansion.
Locally we do not expect this to be the case.
The cumulative frequency-shift and $\tau_{\rm max}$ differ depending on 
the scattering direction. According to Eq.(3), the increment of the
frequency shift of the scattering photon is represented by  
\begin{equation}
 \tilde{\nu}_{i+1} = \tilde{\nu}_i 
  + \psi \left(\varepsilon, w_i \right) \tilde{l}.
\end{equation}
 \end{enumerate}

\section{Results} 

We first compare the results of our 3D simulation technique and improved
approximation to the Ly$\alpha$ opacity to those of LR99.
We then expand upon that previous work by considering the aspherical
expansion of the IGM.
Recently Spergel et al. (2003) have used WMAP data to determine more
precise values for the cosmological model parameters. 
WMAP data also suggest that the epoch of reionization may be at a
redshift larger than 10 (Kogut et al. 2003).
Thus we conclude with a discussion of the effect of redshift and
the cosmological model parameters on the observable Ly$\alpha$ line
profile. 

All figures presenting the results are made by counting the number of
escape photons in the same interval of normalized frequency shift of
$\tilde{\nu}$ in log scale. 
We choose the interval of $10^{-3}$ from $\log{\tilde{\nu}}=-1.0$ to
$2.0$. 
For a clear presentation of the effect of the aspherical expansion,
we count the photons all over the complete solid angle of 4$\pi$. 

 \subsection{Geometrical effect}
In Fig.2 the geometrical effect of using our 3D numerical method for the
scattering is compared to the 2D treatment of LR99. 
The same scattering coefficient used in LR99 is adopted  with $
\varepsilon = 1$. 
The horizontal axis is the normalized frequency, $\tilde{\nu}$, and the
vertical axis is the escape fraction of test photons.
The 2D treatment of scattering by LR99 is the solid line; while our 3D
treatment of the scattering is the dashed line.  
We observe no difference between the lines.
This agreement comes from a simple approximation that the neutral
IGM around the Ly$\alpha$ source has spherical symmetry.
Furthermore, it shows our 3D scheme has sufficient resolution to compare
directly with the results of LR99.
Hereafter, we compare our results with the spherically symmetric case in
3D coordinates, which is called {\it 3D LR99}.
  \begin{figure}[htbp]
   \begin{center}
    \includegraphics{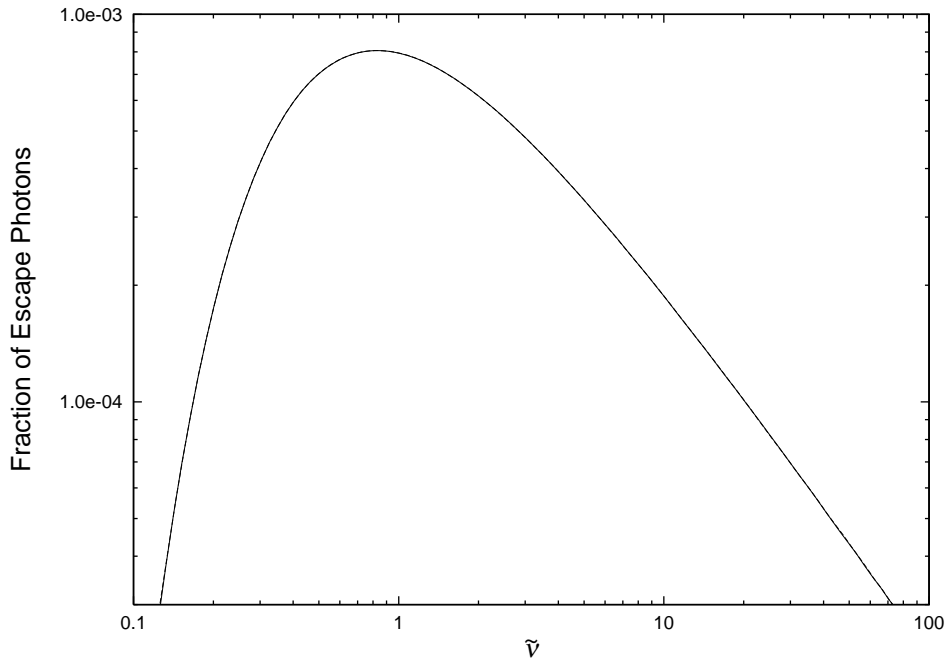}
    \caption{Confirmation of our scheme. Solid line is the spatially 2D
    scattering model and dashed line is the spatially 3D scattering
    model.}  
    \label{fig1}
   \end{center}
  \end{figure}

 \subsection{Opacity evaluation effect}
  \begin{figure}[htbp]
   \begin{center}
    \includegraphics{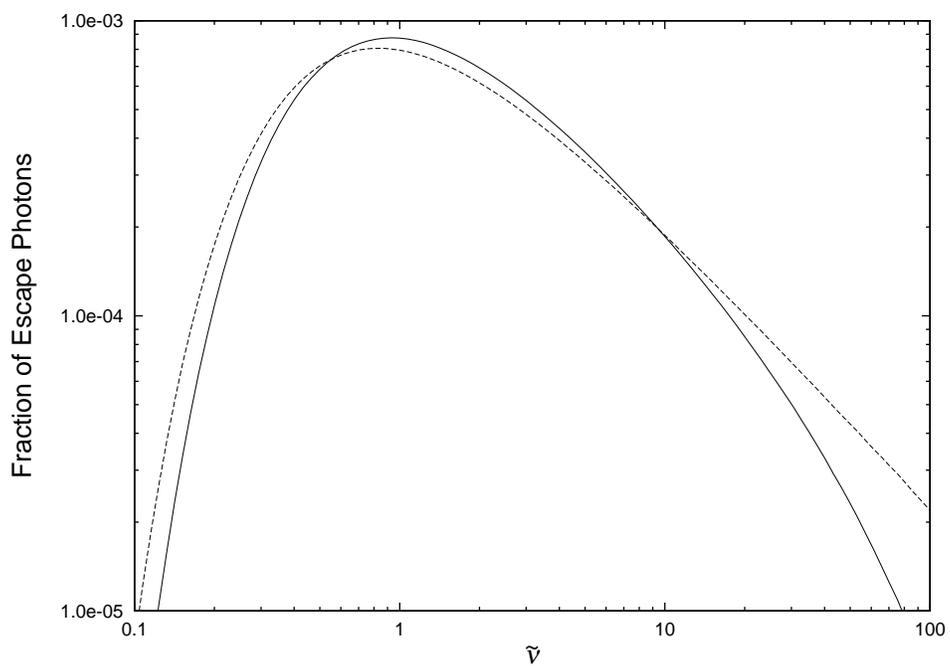}
    \caption{The effect of the selection of opacity is depicted.
    Fully formulated opacity is presented as the solid line ($\varepsilon =1$) 
    and 3D LR99 is the dashed line.} 
    \label{fig2}
   \end{center}
  \end{figure}
How about the effect of opacity?
Fig.{\ref{fig2}} shows the results with the fully formulated opacity and
$\varepsilon = 1$ ({\it solid line}) and that of 3D LR99 ({\it dashed
line}), using the same definition of both coordinates as in figure
{\ref{fig1}}. 
It is found that there are almost no difference between the line
profiles of ours and LR99's.
This result confirms that the Ly$\alpha$ opacity is approximated well in
LR99. 

\begin{figure}[htbp]
   \begin{center}
    \includegraphics{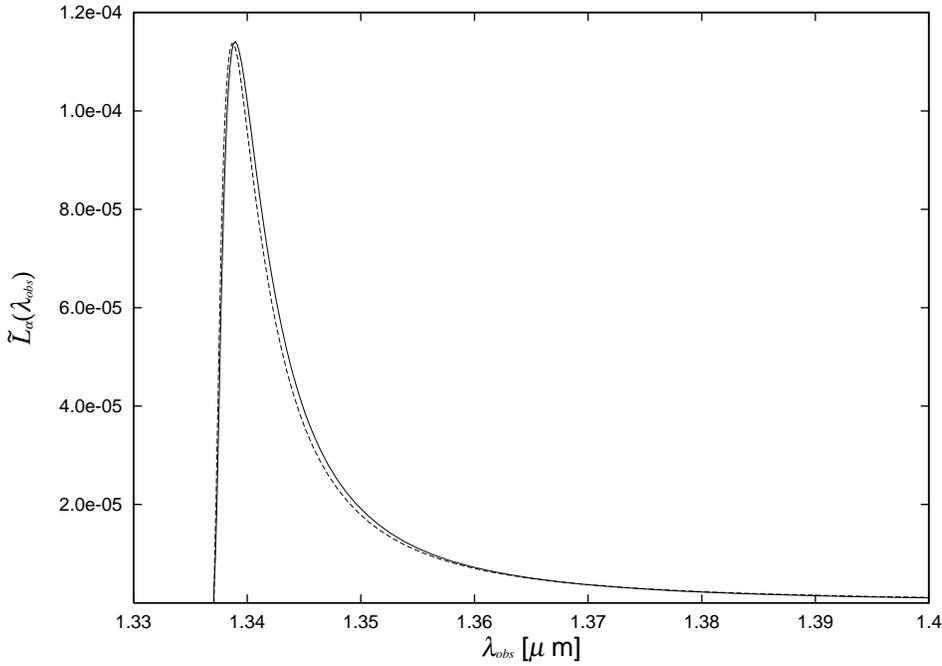}
    \caption{Normalized luminosities of the Ly$\alpha$ line profile as a
    function of observational wavelength $\lambda_{\rm obs}$ [$\mu$m]
    are depicted. 
    Our result is solid line ($\varepsilon =1$) and 3D LR99 is 
    dashed line.}
    \label{fig4}
   \end{center}
  \end{figure}

Transforming $\tilde{\nu}$ to the observational wavelength of
$\lambda_{\rm obs}$ by 
  \begin{equation}
   \lambda_{\rm obs}=\left(1+z_s\right)\lambda_{\alpha}
    \left[1 - \left(\frac{\nu_{\ast}}{\nu_{\alpha}}\right)
     \tilde{\nu} \right]^{-1}{\label{eq-lambda-obs}}, 
  \end{equation}
we compare our model with 3D LR99. 
In Eq.({\ref{eq-lambda-obs}}), the term $\left[1 - \left(\nu_{\ast} /
\nu_{\alpha}\right)\tilde{\nu} \right]^{-1}$ represents the effect of
the cumulative frequency-shift.  
The bin of the wavelength for photon counts corresponds to that of the
frequency. 
We show the normalized luminosity of the Ly$\alpha$ line per unit
wavelength, $\tilde{L}_{\alpha} (\lambda)$, in Fig.{\ref{fig4}}.
This can be converted into the observational $L_{\alpha} (\lambda)$
by multiplying the steady emission rate of Ly$\alpha$ photons at the
source, $\dot{N_{\alpha}}$ (in photons $s^{-1}$), by the Ly$\alpha$
photon energy, $h\nu_{\alpha}$.  
The observed Ly$\alpha$ wavelength emitted at $z_s=10$ is $\lambda_{\rm
obs} = 1.337\ \mu {\rm m}$.
As shown in the figure, the peaks of the line profiles shift to
$\lambda_{\rm peak}\sim 1.339\ \mu {\rm m}$. 
This shift from the line center is $\sim 2\ {\rm nm}$.
We need a wavelength resolution of $R\sim 7.8\times 10^2$ if it is to be
detected. 
The full widths at half maximum (FWHM) of our line profile and 3D LR99
are $5.27\times 10^{-3}\ {\rm \mu m}$ and $4.94\times 10^{-3}\ {\rm \mu
m}$, respectively. 

 \subsection{Aspherical expansion effect}
\begin{figure}[htbp]
   \begin{center}
    \includegraphics{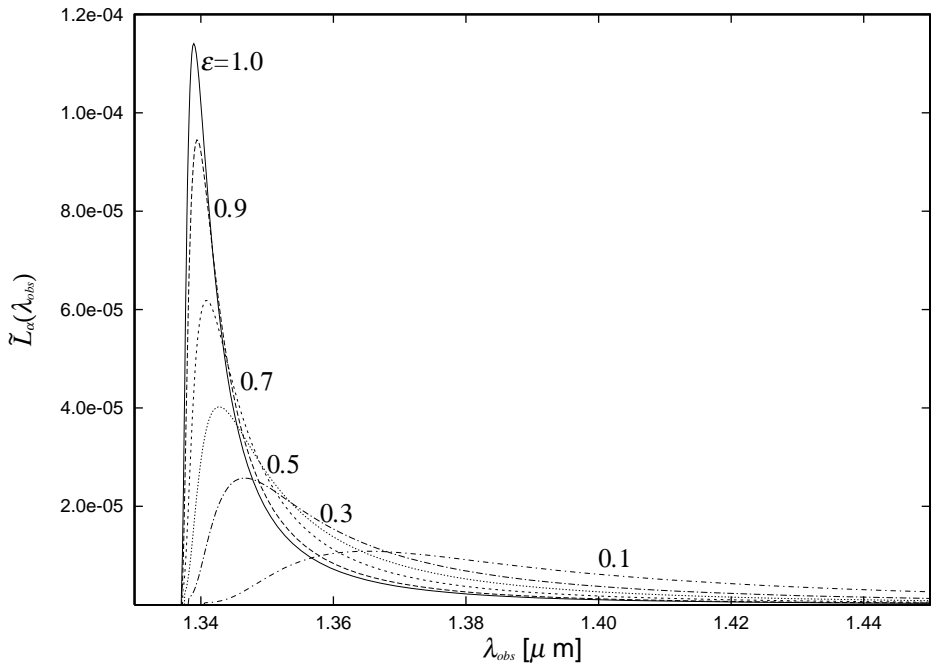}
    \caption{The effect of the aspherical expansion is presented. Each of
    the lines shows 
    $\varepsilon = 1.0$ ({\it solid line}; 
    $\varepsilon = 0.9$ ({\it long dashed line}), 
    $\varepsilon = 0.7$ ({\it short dashed line}), 
    $\varepsilon = 0.5$ ({\it dotted line}), 
    $\varepsilon = 0.3$ ({\it long dash-dotted line}), 
    and 
    $\varepsilon = 0.1$ ({\it short dash-dotted line}), respectively.} 
    \label{fig5}
   \end{center}
  \end{figure}

In this section the effect of the aspherical expansion of the HI IGM is
summarized.  
Fig.{\ref{fig5}} presents the normalized luminosities of Ly$\alpha$ line
profiles for various values of the parameter $\varepsilon$ that
characterizes the asphericity of the IGM. 
Here, the solid line is $\varepsilon = 1.0$, long dashed line is
$\varepsilon = 0.9$, short dashed line is $\varepsilon = 0.7$, dotted
line is $\varepsilon = 0.5$, long dash-dotted line is $\varepsilon =
0.3$, and short dash-dotted line is $\varepsilon = 0.1$.
What is the distinct property of the line profile with $\varepsilon \ne 1.0$? 
The characteristic quantities are summarized in Table {\ref{tab1}},
where some results of other $\varepsilon$'s are also presented. 
The last column in Table {\ref{tab1}} is the necessary resolution to
distinguish  $\lambda_{\rm peak}$ from the Ly$\alpha$ line center of
$(1+z_s)\lambda_{\alpha}$.

  \begin{table}[htbp]
   \begin{center}
    \begin{tabular}[c]{ccccc}
     \hline
     $\varepsilon$ & FWHM & $\lambda_{\rm peak}$ & peak value & $R$\\
     \hline
     1.0 & $5.27\times 10^{-3}\ {\rm \mu m}$ & $1.3390\ {\rm \mu m}$
     & $1.14\times 10^{-4}$ & $7.8\times 10^2$ \\
     0.9 & $6.38\times 10^{-3}\ {\rm \mu m}$ & $1.3394\ {\rm \mu m}$
     & $9.45\times 10^{-5}$ & $6.2\times 10^2$ \\
      0.8 & $7.86\times 10^{-3}\ {\rm \mu m}$ & $1.340\ {\rm \mu m}$
      & $7.69\times 10^{-5}$ & $4.7\times 10^2$ \\
     0.7 & $9.77\times 10^{-3}\ {\rm \mu m}$ & $1.341\ {\rm \mu m}$
     & $6.19\times 10^{-5}$ & $3.7\times 10^2$ \\
      0.6 & $1.21\times 10^{-2}\ {\rm \mu m}$ & $1.342\ {\rm \mu m}$
      & $4.97\times 10^{-5}$ & $3.0\times 10^2$ \\
     0.5 & $1.48\times 10^{-2}\ {\rm \mu m}$ & $1.343\ {\rm \mu m}$
     & $4.03\times 10^{-5}$ & $2.5\times 10^2$ \\
      0.4 & $1.80\times 10^{-2}\ {\rm \mu m}$ & $1.344\ {\rm \mu m}$
      & $3.27\times 10^{-5}$ & $2.0\times 10^2$ \\
     0.3 & $2.30\times 10^{-2}\ {\rm \mu m}$ & $1.346\ {\rm \mu m}$
     & $2.57\times 10^{-5}$ & $1.5\times 10^2$ \\
      0.2 & $3.25\times 10^{-2}\ {\rm \mu m}$ & $1.352\ {\rm \mu m}$
      & $1.86\times 10^{-5}$ & $94$ \\
     0.1 & $1.11\times 10^{-2}\ {\rm \mu m}$ & $1.365\ {\rm \mu m}$
     & $1.09\times 10^{-5}$ & $48$ \\ 
     \hline
    \end{tabular}
    \caption{Summary of the effect of the aspherical deceleration.} 
    \label{tab1}
   \end{center}
  \end{table}

According to table 1, the distinct effects of the asphericity are the
shift of the peak and the broadening of the FWHM of the line profiles.
These are clearly seen in the line profile with $\varepsilon = 0.1$.
If there is strong asphericity of the expanding HI IGM, we will detect
broadening of FWHM and the shift of the peak wavelength of the line
profile.  
This could be observable with the JWST and signal the presence of the LR
halo. 

For increasing asphericity, i.e. a decreasing $\varepsilon$, the optical
depth for photons to scatter becomes larger than in the spherical model.
Thus, although the typical shift per scattering due to the expansion
of the universe becomes small, the number of scattering events
increases, thus increasing the probability that a photon will be shifted
far from the line center.

 \subsection{The Effect of Redshift and Cosmological Parameters}
\begin{figure}[htbp]
  \begin{center}
   \includegraphics{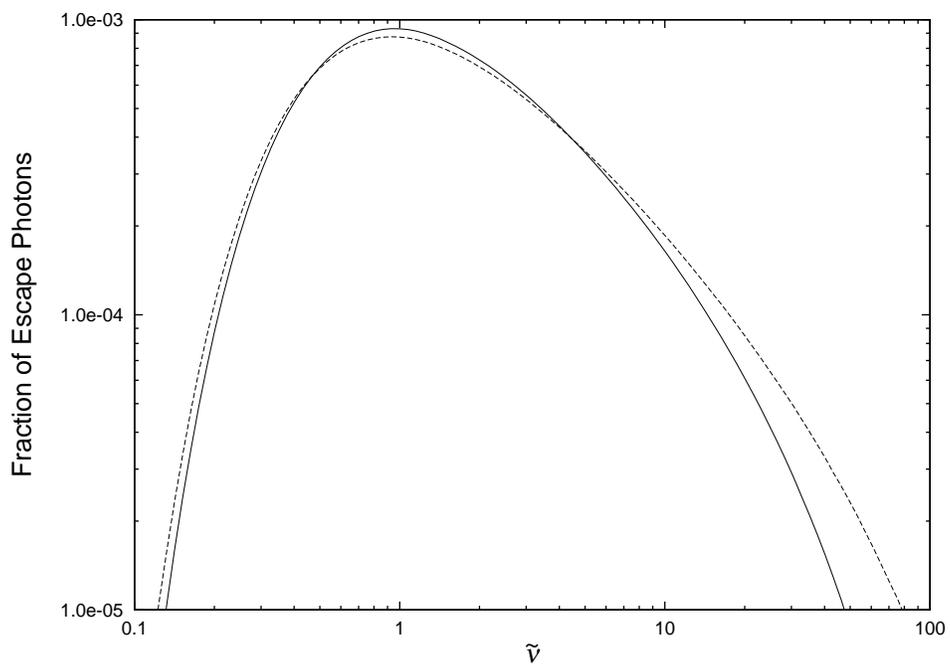}
   \caption{Line profiles of $z_s$ = 20 ({\it solid line}) and $z_s$
   = 10 ({\it dashed line}) with  $\varepsilon =1$ are presented.}
   \label{fig:z=20_1}
  \end{center}
\end{figure}

WMAP data suggest that cosmic reionization may take place at $z_s \sim
20$ (Kogut et al. 2003).
In Fig.{\ref{fig:z=20_1}} we show the results of Ly$\alpha$ line
transfer on the Ly$\alpha$ line profiles from a source at $z_s=
20$ (solid line) compared to a source at $z_s=10$ (dashed line).
The coordinates are the same as in Fig.{\ref{fig2}}.
We find the line profile of $z_s = 20$ coincides with that of $z_s =10$
very well.
This comes from the fact that most of the Ly$\alpha$ photons can escape
from the IGM when their cumulative frequency shift becomes $\sim
\nu_{\ast}$, and it takes place at the distance of $\sim r_{\ast}$
from the source.

\begin{figure}[htbp]
  \begin{center}
   \includegraphics{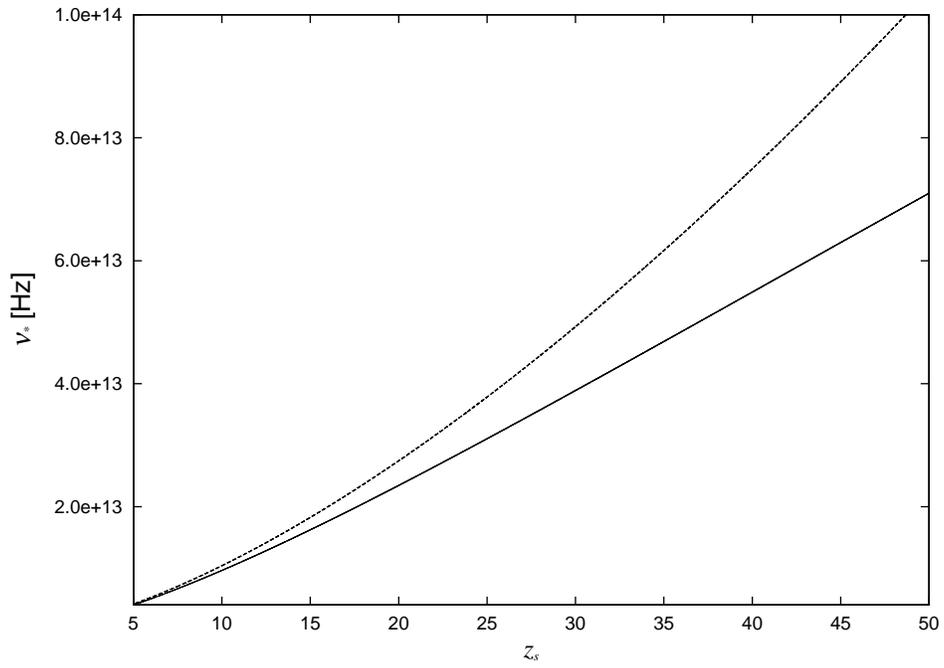}
   \caption{Redshift dependence of the characteristic frequency shifts
   is depicted. Solid line is $\nu_{\ast}$, and dashed line is
   $\nu_{\ast}^{\rm app}$.} 
   \label{z_nu}
  \end{center}
\end{figure}

While $\nu_{\ast}$ and $r_{\ast}$ depend upon redshift, the shape of the
two line profiles as functions of the normalized $\tilde{\nu}$ are very
similar. 
The effect of redshift is not significant except in the normalizations.
This result presents the useful implication that when we examine line
profiles at a variety of redshifts, it is sufficient for us to know the
characteristic frequency shift and radius at each redshift.
Hence, we  summarize the redshift dependence of $\nu_{\ast}$ in 
Fig.{\ref{z_nu}}.
Here, the solid line is $\nu_{\ast}$ of Eq.(10), while 
the dashed line is  $\nu_{\ast}^{\rm app}$ which  
is adopted in LR99 and determined by Eq.(6) and Eq.(9).
For simplicity, we present the cases with $\varepsilon =1$.
We hope these results are useful for the heterorelation of future
observations. 

\begin{figure}[htbp]
  \begin{center}
   \includegraphics{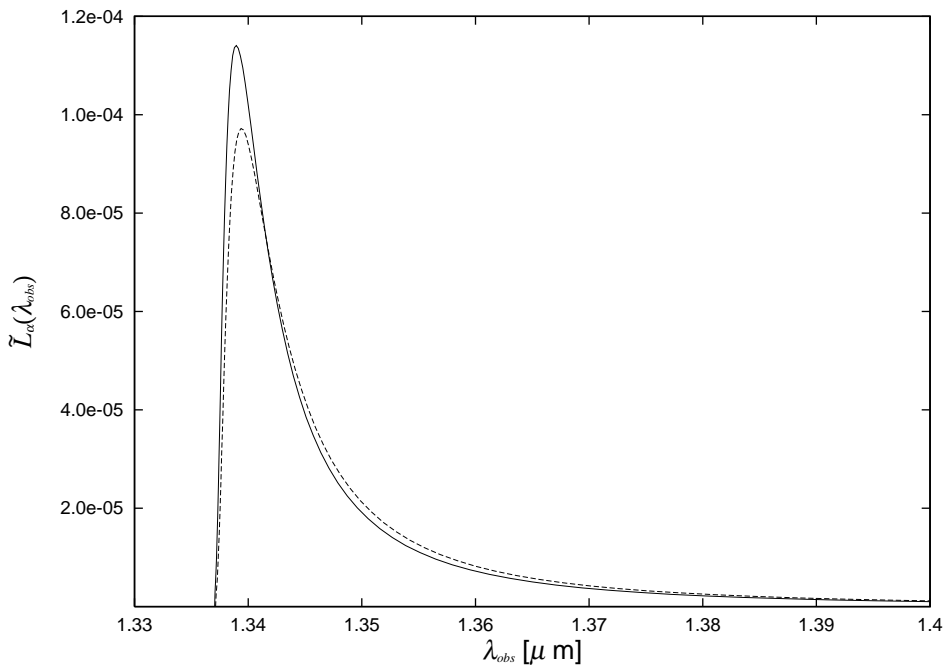}
   \caption{Observational line profiles at $z_s$ = 10 are presented.
    Solid line is the model with LR99 cosmological parameters,
    while dotted line is those of WMAP.}
   \label{map}
  \end{center}
\end{figure}

We also simulate the line transfer with the cosmological parameters
found by Spergel et al. (2003) in their analysis of WMAP data, who
suggest $h_0 = 0.71^{+0.04}_{-0.03}$, $\Omega_{\rm M} h_0^2 = 0.135
^{+0.008}_{-0.009}$, $\Omega_{\rm b} h_0^2 = 0.0224 ^{+0.009}_{-0.009}$,
and $\Omega _\Lambda = 1.0 ^{+0.02}_{-0.02} - \Omega_{\rm M}$.
Fig.{\ref{map}} presents the results at $z_s=10$.
The solid line adopts the  cosmological parameters of LR99,
while the dashed line uses those of  Spergel et al. (2003).
We find the effect of the cosmological parameters on the line profile is
small.

\section{SUMMARY}

In this work we extended LR99 by improving the numerical treatments of
the Ly$\alpha$ opacity and scattering processes. 
As part of these improvements, the effect of asphericity in the
expanding IGM on the Ly$\alpha$ profiles was examined.
The results confirm LR99 very well when the aspherical deceleration is
small. 
The effect of aspherical expansion on the line profile is reflected in a
broadening of the FWHM and a shift of the peak intensity to longer
wavelengths over that in the spherical model.
The effect of the redshift of reionization is reflected in the
normalization of the frequency shift and escape distance from the source
rather than in the shape of the normalized line ratio.
Our results were insensitive to small changes in the cosmological
parameters, with little difference in the results found using the
cosmological model of LR99 and that derived recently from WMAP data.
To detect both properties will be important to find the initial stage of
the formation of LSS.  
Fortunately, it requires a spectroscopic resolution of $R\sim 100-1000$,
which is within the design characteristics for future observational
projects like JWST.


\acknowledgments

We are grateful to the referee for his/her advisable comment and
excellent refereeing, which improve our paper very much in both
content and presentation.  We thank Profs. S. Inagaki, S. Mineshige,
T. Totani for their continuous encouragement.



\begin{thebibliography}{}

\bibitem[Abel, Bryan, \& Norman (2002)]{abn00}
Abel, T., Bryan, G. L., \& Norman, M. L. 2000, ApJ, 540, 39

\bibitem[Barkana \& Loeb (1999)]{bl99}
Barkana, R. \& Loeb, A. 1999, ApJ, 523, 54

\bibitem[Breit \& Teller (1940)]{bt40}
Breit, G., \& Teller, E. 1940, ApJ, 91, 215

\bibitem[Dey et al. (1998)]{detal98}
Dey, A., Spinrad, H., Stern, D., Graham, J. R., \& Chaffee, F. H. 1998,
ApJ, 498, L93

\bibitem[Gnedin \& Ostriker (1997)]{go97}
Gnedin, N. Y. \& Ostriker, J. P., 1997, ApJ, 486, 581

\bibitem[Gunn \& Peterson (1965)]{gp65}
Gunn, J. E., \& Peterson, B. A. 1965, ApJ, 142, 1633

\bibitem[Hu, Cowie, \& McMahon (1998)]{hcm98}
Hu, E. M., Cowie, L. L., \& McMahon, R. G. 1998, ApJ, 502, L99

\bibitem[Kamaya \& Silk (2002)]{ks02}
Kamaya, H., Silk, J., 2002, MNRAS 332, 251

\bibitem[Kogut et al.(2003)]{kog03}
Kogut, A., et al. 2003, ApJS, 148, 161

\bibitem[Lin et al. (1965)]{lms65}
Lin, C. C., Mestel, L., \& Shu, F. H., 1965, ApJ, 142, 1431


\bibitem[Loeb \& Rybicki (1999)]{lr99}
Loeb, A., \& Rybicki, G. B. 1999, ApJ, 524, 527 (LR99)


\bibitem[Miralda-Escud\'{e}, Haehnelt, \& Rees (2000)]{mhr00}
Miralda-Escud\'{e}, J., Haehnelt, M. \& Rees, M. J. 2000, ApJ, 530, 1


\bibitem[Shchekinov (1991)]{siu91}
Shchekinov, Iu., 1991, Ap\&SS, 175, 57

\bibitem[Spergel et al. (2003)]{wmap03}
Spergel, D. N., et al., 2003, ApJS, 148, 175

\bibitem[Spinrad et al. (1998)]{spin98} 
Spinrad, H., Stern, D., Bunker, A., Dey, A., Lanzetta, K., Yahil, A., 
Pascarelle, S., \& Fern\'{a}ndez-Soto, A. 1998, AJ, 116, 2617

\bibitem[Spitzer \& Greenstein (1951)]{sg51}
Spitzer, L. J., \& Greenstein, J. L., 1951, ApJ, 114, 407

\bibitem[Weymann et al. (1998)]{wey98}
Weymann, R. J., Stern, D., Bunker, A., Spinrad, H., Chaffee, F. H.,
Thompson, R. I., \& Storrie-Lombardi, L. J. 1998, ApJ, 505, L95

\bibitem[Witt (1977)]{witt77}
Witt, A. N. 1977, ApJS, 35, 1
\end{thebibliography}
\end{document}